\documentclass[11pt,a4paper]{article}
\usepackage{amsmath}
\usepackage{graphicx}
\usepackage{bm}
\usepackage[T1]{fontenc}
\usepackage[utf8]{inputenc}
\usepackage{authblk}
\usepackage{wrapfig}
\usepackage{authblk}
\usepackage{enumitem}
\usepackage{caption}
\usepackage[numbers,sort&compress]{natbib}

\usepackage{colortbl}
\usepackage{tabularx}

\usepackage[colorlinks, urlcolor=blue, linkcolor=blue]{hyperref}

\newcommand*{\email}[1]{%
    \normalsize\href{mailto:#1}{#1}\par
    }

\title{\vspace{-3cm}FNAL PIP-II Accumulator Ring}
\date{\vspace{-5ex}}
\author[1]{ William Pellico\thanks{\email{pellico@fnal.gov}}}
\author[1]{Chandra Bhat}
\author[1]{Jeffrey Eldred}
\author[1]{Carol Johnstone}
\author[1]{John Johnstone} 
\author[1]{Kiyomi Seiya}
\author[1]{Cheng-Yang Tan}
\author[1]{Matthew Toups}
\author[2]{Patrick deNiverville}
\author[2]{Richard Van De Water}

\affil[1]{Fermi National Accelerator Laboratory}

\affil[2]{Los Alamos National Laboratory}

\begin{document}
\maketitle
\begin{abstract}
The FNAL accelerator complex is poised to reach MW neutrino beams on target for the exploration of the dark sector physics and rare physics program spaces. Future operations of the complex will include CW linac operations at beam intensities that have not been seen before~\cite{PIP2,RCS_LOI}.  The ambitious beam program relies on multi-turn H$^{-}$ injection into the FNAL Booster and then extracted into delivery rings or the Booster Neutrino Beam (BNB) 8 GeV HEP program. A new rapid-cycling synchrotron (RCS) will be required to reach the LBNF goal of 2.4 MW because of intense space-charge limitations. There are many accelerator engineering challenges that are already known and many that will be discovered.  This proposal calls for an intermediate step that will both facilitate the operation of Booster in the PIP-II era and gain operational experience associated with high power injection rings. This step includes the design, construction and installation of a 0.8 GeV accumulator ring (upgradeable to 1+ GeV) to be located in the PIP-II Booster Transfer Line (BTL).  The PIP-II accumulator ring (PAR) may be primarily designed around permanent magnets or use standard iron core magnet technology with an aperture selected to accommodate the desired high intensity protons at 0.8 GeV.
\end{abstract}

\section{Design Concept}

This paper highlights just a few of the critical areas of operations that an accumulator ring will address and how it can make an impact to the Fermilab HEP program within this decade. The success of the Fermilab upgrade plans will be ensured with this proposed PIP-II Accumulator Ring (PAR). The design of PAR will be guided by these operational needs.
A previous proposed accumulator ring that was to be located on the Booster outer tunnel wall, where there is sufficient space, was provided in a 2021 Snowmass LOI. 
This updated paper will describe PIP-II injection into PAR which will be located in a largely open field area adjacent to the old Booster beam dump. Similar to the previous LOI, this paper contains much of the same needs and benefits.

\begin{wrapfigure}{C}{0.45\textwidth}
\centering
\includegraphics[width=0.45\textwidth]{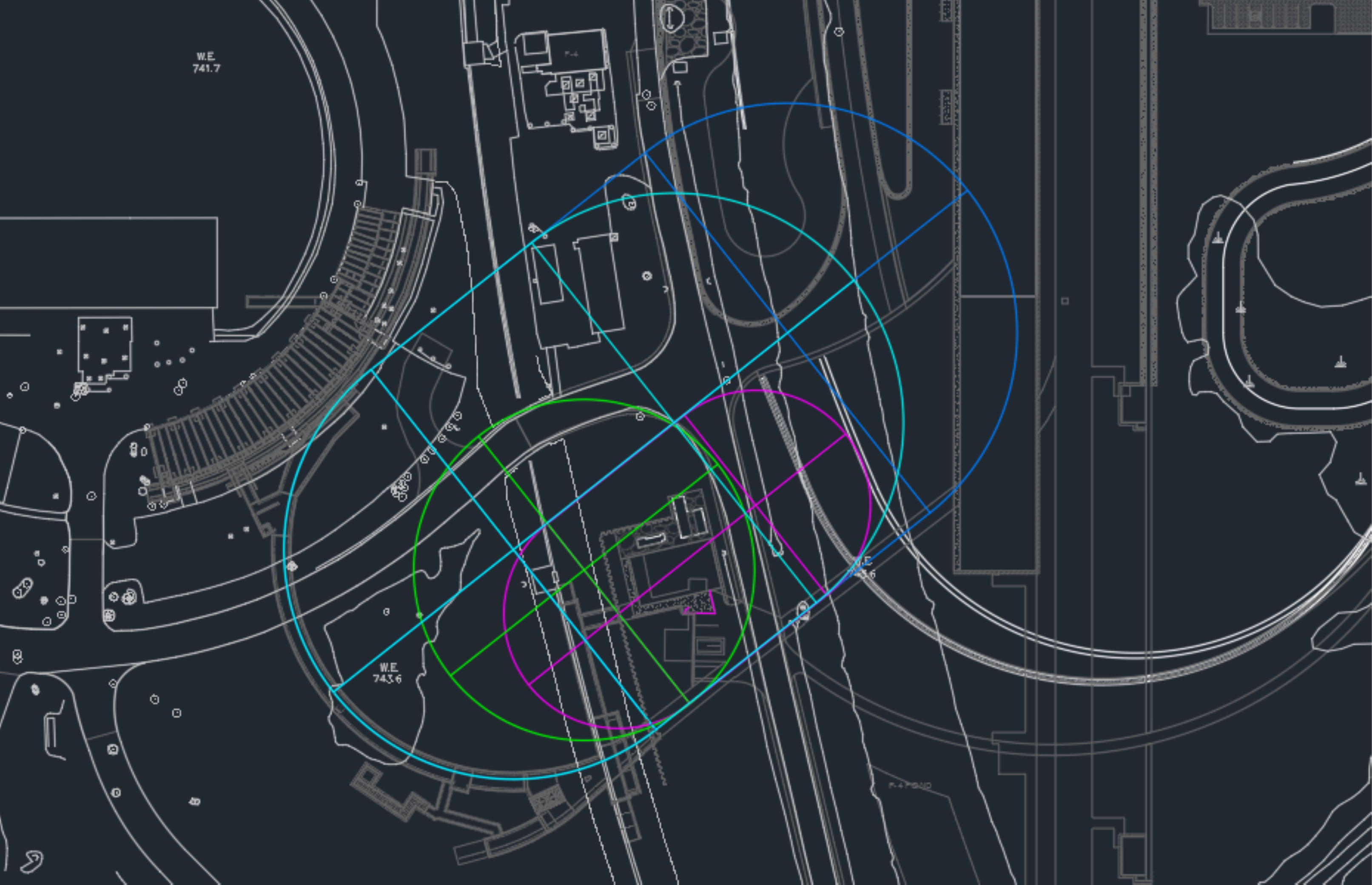}
\caption{Ring options in the PIP-II BTL footprint area. The ring drawn in green is the approximate shape of PAR. The exact location of PAR along the BTL is still to be determined.}
\label{fig:footprintlayout}
\end{wrapfigure}

The initial investigation for staging of PAR included:
\begin{enumerate}
\item impact to PIP-II construction.
\item alignment to PIP-II beam operations.
\item existing utility infrastructure:
\begin{enumerate}
\item MR tunnel enclosure
\item Service buildings
\end{enumerate}
\item potential physics program implementation.
\end{enumerate}

\begin{wrapfigure}{C}{0.45\textwidth}
\centering
\includegraphics[width=0.45\textwidth]{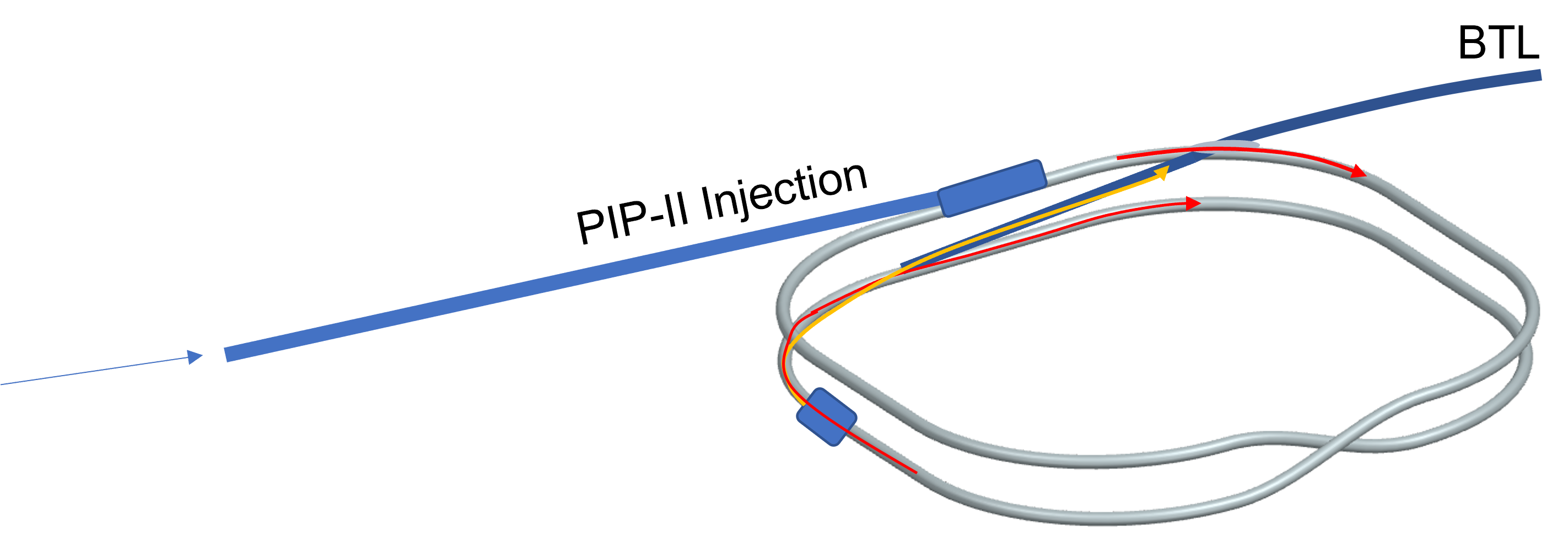}
\caption{Folded accumulator ring layout aligned with the PIP-II BTL.}
\label{fig:Footprint}
\end{wrapfigure}

Several rings sizes and shapes were considered that are shown in Fig.~\ref{fig:footprintlayout}. The chosen solution is a tunnel that is half the circumference of Booster which contains a folded ring shown in Fig.~\ref{fig:Footprint}. The total path length of the folded ring is exactly the Booster circumference. Table~\ref{tab::params} summarizes the basic PAR parameters.

\begin{table}
\centering
\begin{center}
\caption{PAR parameters used in this proposal design.} 
\label{tab::params}
\begin{tabular}{|c |c |c|c|c|}
\hline
{\bf Accelerator } & \bf Aperture & \bf Total Ring Cir & \bf Tunnel Cir & \bf Cycle Rate \\
\hline
Accumulator Ring & $\sim\!3${\tt"}  &  $\sim\!477$~m & $\sim
\!240$~m & $\sim\!120$ Hz \\
\hline
\end{tabular}
\end{center}
\end{table}

Figure~\ref{fig:footprintlayout} shows the proposed location of PAR (green) in a largely open space that is adjacent to Booster. This location will allow for some flexibility in the lattice design. PAR will need to match the PIP-II Booster Transfer Line (BTL) at a long straight. The PAR injection region will be designed to both reduce beam loss and to absorb the waste and scattered beam from the foil. Additional required components, like the RF cavities and extraction hardware, can be located at one of the other 4 planned long straight sections.  

The principal motivation for PAR is to supply high intensity proton beams to 0.8 GeV HEP users. The power available for the new extraction line will depend on shielding and hardware pulse limitations. The initial beam power estimate is >100 kW baseline to a Dark Sector (DS) program.

In the configuration where PAR is used to load Booster, the beam is accumulated in PAR to full intensity and then transferred to Booster via single turn injection and then accelerated immediately. The Booster should see reduced injection losses when compared to the current PIP-II injection process because H- stripping is no longer done there. Another benefit is that phase locking of the PAR RF to the PIP-II Linac RF is much simpler than locking to the Booster RF because of the flat field of PAR. This configuration also supports an 8 GeV neutrino HEP program with an estimated 87 kW of beam power.

In summary, PAR will take advantage of Fermilab's significant facility upgrade to open new physics opportunities.

\section{Magnets and Hardware}
At this early stage of the proposal, both standard iron core magnet technology and permanent magnet technology will be explored. However, permanent magnets will be mainly discussed here. Permanent storage rings are being built or have been in use in many accelerator facilities. This is because permanent magnets have the the advantages of zero power consumption, higher reliability, and lower assembly costs when compared to powered magnets.
 The FNAL Recycler, which uses permanent magnets \cite{Volk_2011}, has proven itself to be a critical component for Fermilab's success during the collider physics program and now in the high intensity neutrino and rare decay programs.  The permanent magnet storage ring is an economical platform to manipulate beams for loading into a rapid cycling accelerator because it does not have ramp time constraints.   
 However, for brevity and the expected small cost difference, the arguments for using either permanent magnets or standard magnets will not be included in this paper. The choice of magnet technology will be fully explored and documented in the planning of PAR.

This paper will provide a preliminary concept with some of the parameter space mapped out.   
For evaluation purposes and to help guide lattice efforts, the required magnetic field for the permanent magnets have been set to be around 0.3 T for a $3.1$ inch gap. The magnets that will be investigated will be either Nd2Fe14B or Sm2Co17\cite{Kraus,CERN, Green} because significant community effort have been invested in designs using these materials.  However, the option of using a combination of electro-permanent magnet designs has not been ruled out. This is to allow for PAR to be upgraded to accept an increase in injected beam energy from the PIP-II Linac without adding more magnets.

As noted earlier, PAR consists of one ring that is folded to form two rings with each ring half the size of the Booster circumference. See Fig.~\ref{fig:Footprint}. The folded ring will have to meet challenging lattice requirements because of its small size. The magnet designs that can meet the bending constraints and still fit between the two rings are being investigated. Additionally, early results suggest edge focusing is important for achieving the required zero dispersion in the straight sections and so must be taken into account.

\section{Lattice Design}

The early lattice development has been focused on meeting the requirement that the lattice has zero dispersion at two approximate 9~m long straight sections to accommodate injection and extraction and two 5~m short straights for RF and other components.

The values for tunes and chromaticities are more flexible and will evolve as the lattice matures. The injection region is challenging due to the desire to accommodate beam dumps for the neutrals, H0, and unstripped H- passing through the foil from the injection process. An efficient and controlled H- stripping process will need significant space because of the high repetition rate.  The lattice design also needs to consider the spacing of the magnets in the critical regions: (a) where the rings cross, (b) the injection/extraction region and (c) RF region, so that there will not be any physical interference.

\subsection{Lattice and optics}

PAR is composed of two rings, each of which is exactly half the Booster circumference, with one ring stacked vertically above the other. The lattice in each ring comprises four modular optical constructs:
\begin{enumerate}
\item Arc cells,
\item Dispersion suppressor cells,
\item Two long straight insertion sections, and
\item Two short straight insertion sections.
\end{enumerate}

One pair of the long straights are located in the same horizontal positions, but are offset vertically. One of the straights of this pair is used for beam injected from the BTL onto a stripping foil. The other straight which is in the other ring is used for extraction to Booster. See Fig.~\ref{fig:Footprint}. The second pair of long straights will accommodate the vertical transfer of beams between the two machines, i.e.~the crossing region. The short straights will be used for RF and other components.

\begin{figure}[htb]
\centering
\includegraphics[width=0.80\textwidth]{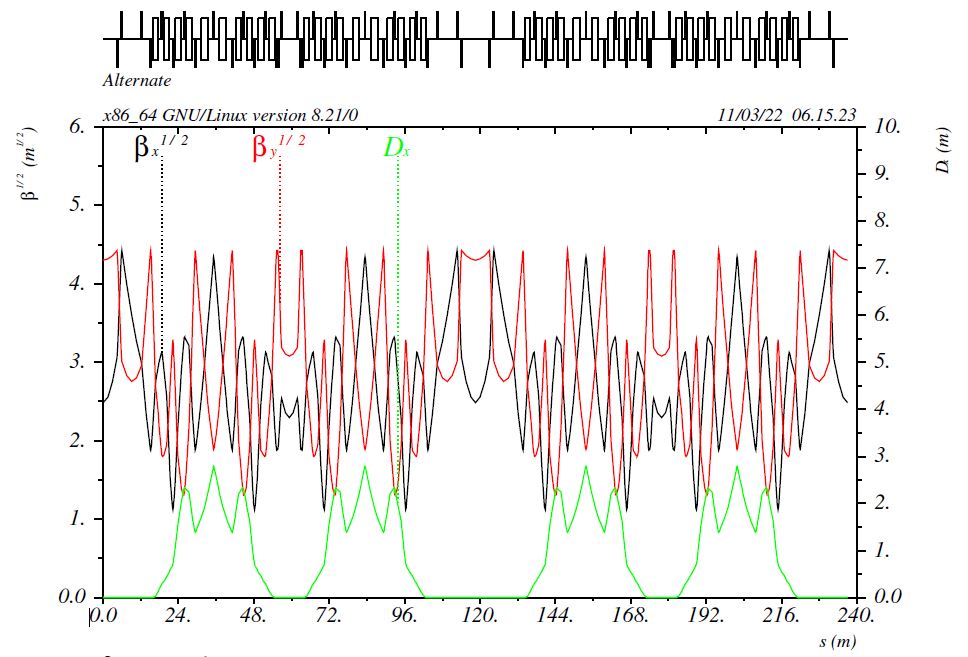}
\caption{A preliminary lattice design for one of the rings of PAR is shown here. The other ring has the same lattice functions.}
\label{fig:lattice}
\end{figure}

The lattice functions of one of the rings (the other ring has the same lattice) are shown in Fig.~\ref{fig:lattice}.  The major ring characteristics are shown in Table~\ref{tab::ringparams}.

\begin{table}
\centering
\begin{center}
\caption{Major characteristics of PAR} 
\label{tab::ringparams}
\resizebox{\textwidth}{!}{
\begin{tabular}{|c |c |c|c|c|c|c|c|c|}
\hline
{\bf Parameter } & \bf Circum. & $\bm{ \mu_x}$ & $\bm{ \mu_y}$ &$\bm{\beta_x}$ \bf max &$\bm{\beta_y}$ \bf max &$\bm{\beta_x}$ \bf foil &$\bm{\beta_y}$ \bf foil &$\bm{\eta_x}$\\
\hline
 \bf Value & $237.1$~m  &  $6.6404$ & $6.0939$  & $19.557$~m & $19.557$~m & $6.166$~m & $18.497$~m & $2.7892$~m\\
\hline
\end{tabular}
}
\end{center}
\end{table}

There are only two kinds of dipoles in the ring but a large variety of quadrupoles. The quadrupole lengths and/or gradients are quite different between the arcs, suppressor and the long and short straights. The design choice becomes whether to have: 
\begin{enumerate}[label=(\alph*)]
\item a large number of separate power supplies and a small number of quad types, or the reverse
\item only a few power supplies and a larger number of quadrupole types. 
\label{item::options}
\end{enumerate}
In the tables that follow, it is option~\ref{item::options} that is largely followed by electing to choose only one supply for F quads and a second for D quads. This results in 10 different quadrupole lengths. Other combinations of supplies and lengths are, of course, possible.

\subsubsection{Arc cells}

The arc cells are chosen to have the same physical length as the cells in the BTL line and have the same $90^\circ$ of betatron phase advance per cell. This choice is not obligatory but will ease in optically matching beam transfers. Each half-cell contains a single dipole and a Booster-style corrector package. Parameters of an arc cell are listed in Table~\ref{tab::arc}.

\begin{table}
\centering
\begin{center}
\caption{Parameters of an arc cell} 
\label{tab::arc}
\resizebox{\textwidth}{!}{
\begin{tabular}{|c |c |c|c|c|c|c|c|}
\hline
{\bf Device } & \bf Cell &\multicolumn{2}{c|}{ \bf Dipoles} & \multicolumn{2}{c|}{ \bf QF} &\multicolumn{2}{c|}{\bf QD} \\
\hline
\bf Param.& \bf length &\bf length &\bf strength & \bf length& \bf gradient &\bf length& \bf gradient \\
\hline
\bf Value & 11.795 m & 1.6764 m & 0.52343 T & 0.21590 m & 5.11671 T/m & 0.21590 m & 5.38490 T/m\\
\hline
\end{tabular}
}
\end{center}
\end{table}

\subsubsection{Dispersion suppressors}

The purpose of the suppressors is two-fold:
\begin{enumerate}
	\item Eliminate dispersion from the straight insertions, and
	\item 	Keep maximum dispersion at a manageable level.
\end{enumerate}

A suppressor unit comprises of two cells of $90^\circ$ phase advance each. This then guarantees that the entrance and exit values of $\beta$’s, $\alpha$’s are equal. Since the 8 dispersion suppressor units comprise the bulk of the lattice, these suppressor cells largely define the ring characteristics.

The suppressor cell lengths and bends are chosen to satisfy the constraint
\begin{equation}
    \ell_{\rm DS}\theta_{\rm DS} = \frac{1}{2}\ell_{\rm ARC}\theta_{\rm ARC}
\label{eq::arc1}
\end{equation}
where $\ell_{\rm DS}$ and $\theta_{\rm DS}$ are the length and bend angle of the dispersion suppressors respectively. And $\ell_{\rm ARC}$ and $\theta_{\rm ARC}$ are the length and bend angle of the arc cells respectively.

The suppressor parameterization described in Eq.~\ref{eq::arc1} optimizes the use of space and maximizes the free space available for the straight sections. Each dispersion suppressor half-cell contains one dipole and a Booster-style corrector. The parameters of the suppressor cell are listed in Table~\ref{tab::ds}. 

\begin{table}
\centering
\begin{center}
\caption{Parameters of an arc cell} 
\label{tab::ds}
\resizebox{\textwidth}{!}{
\begin{tabular}{|c |c |c|c|c|c|c|c|c|c|}
\hline
{\bf Device } & \bf Cell &\multicolumn{2}{c|}{ \bf Dipoles} & \multicolumn{2}{c|}{ \bf QF} &\multicolumn{2}{c|}{\bf QD} &\multicolumn{2}{c|}{\bm{${\rm QD}_{\rm DS}-{\rm QD}_{\rm ARC}$}}\\
&  &\multicolumn{2}{c|}{} & \multicolumn{2}{c|}{} &\multicolumn{2}{c|}{} &\multicolumn{2}{c|}{\bf interface}\\
\hline
\bf Param.& \bf length &\bf length &\bf strength & \bf length& \bf gradient &\bf length& \bf gradient & \bf length & \bf gradient\\
\hline
\bf Value & 7.077 m & 1.3970 m & 0.52343 T & 0.18558 m & 5.11671 T/m & 0.51350 m & 5.38490 T/m & 0.36470 m & 5.38490 T/m\\
\hline
\end{tabular}
}
\end{center}
\end{table}

The prescription described in Eq.~\ref{eq::arc1} works exceptionally well in the Fermilab Main Injector (MI) and results in 2 dipole lengths and just 3 quadrupole lengths. Unfortunately, when compared to MI, it is not as simple here because of the much smaller size of the PAR rings which require the bending per suppressor cell to be very large --- $\sim\!300$~mrad. This is $\sim\!10\times$ that of MI. Dipole focusing is a major contributor to the optics and not a second order effect, as it is in MI, and cannot be ignored.

To achieve a perfect optical match between a suppressor unit and an arc cell while also killing dispersion exactly, it was found necessary to introduce large edge focusing ($-51.2^\circ$) into the suppressor dipoles for compensation.  This edge focusing also results in dispersion suppression quadrupole strengths and lengths that do not obey the simple thin lens rule that the application of Eq.~\ref{eq::arc1} would predict.

\subsubsection{Tune and transverse envelope control}

Three conventional methods exist for controlling the beam envelope and phase advance using linear fields, i.e.~dipole and quadrupole fields.\footnote{Here we refer to the linear lattice functions and the linear tune.} One is the weak focusing principle used in classical cyclotrons in which changes in path length through the magnetic field as a function of transverse position focuses the beam, but only in the bend plane (which is typically horizontal).  Vertical control is achieved by either an angular rather than normal crossing of the pole tip edge (as in a pure sector magnet) or radial shaping of the pole-tip as in the Azimuthally Varying Field (AVF) cyclotrons or spiral sector Fixed Field Gradient Accelerators (FFGA) which presents a varying edge angle increasing with radius and therefore energy.  Edge effects are almost always weaker than focusing from path length differences from a pure dipole field but can be applied to supplement the strong focusing from linear gradient fields which will be described next. 

By applying an edge angle in a dipole and an alternating gradient in quadrupoles, two focusing techniques can be applied:  edge focusing which occurs in the fringe fields of a dipole and an alternating, strong focusing gradient, or quadrupole, focusing which is the main technique used in a synchrotron.  Strong focusing techniques are capable of focusing equally in both planes with much stronger focusing strength, particularly in the case of the quadrupole field. As is well known, stronger focusing results in larger phase advances, shorter focal lengths, and correspondingly, higher machine tunes than achievable in weak-focusing machines, i.e.~stronger envelope control.

\subsubsection{General principles}

When a vertically oriented (horizontally bending) dipole field is present, the physical magnet edge angle brings with it a horizontally focusing or defocusing effect, or no effect in the case of a rectangular magnet – no amplitude-dependent path length differences exist from the body field. Weak focusing by the dipole field in the body of the magnet does not affect the vertical plane. However, focusing by the fringe field of the magnet depends on the angle through which the beam traverses the fringe field. This effect is essentially equivalent to a quadrupole located at each magnet edge:  it can be either focusing horizontally and defocusing vertically, or the reverse. Note: a normal entrance angle has no focusing effect.

\subsubsection{Thin lens formalism}

Linear dynamics can be expressed and understood in terms of the three “conventional” transverse focusing principles outlined above.  To understand the interplay between strong, weak and edge focusing, a simple linear, thin-lens matrix model serves as a guiding example. The approach is most easily explained using sector magnets, then adding a gradient field and edge angles.  The following first order matrix is used for a horizontally focusing sector magnet with an edge angle
\begin{equation}
M=
\begin{pmatrix}
1 & 0\\
-\tan\eta/\rho_0 & 1
\end{pmatrix}
\begin{pmatrix}
\cos\Theta & \frac{1}{\sqrt{K}}\sin\Theta\\
-\sqrt{K}\sin\Theta & \cos\Theta
\end{pmatrix}
\label{eq::gp1}
\end{equation}
where $\Theta = \sqrt{K}\ell$ and $K = k_0 + 1/\rho_0^2$ for a CF sector magnet. If one adds an edge angle, where one adopts the sign convention: $\eta>0$ is outward, or away from the body of the magnet and thus increases net horizontal focusing in all magnets.  Reducing to the thin lens formalism, the matrices starting from the center of the quadrupole magnet in a half FODO cell through its edge to lowest order are:
\begin{equation}
\begin{aligned}
  \begin{pmatrix}
1 & 0\\
-\tan\eta/\rho & 1
\end{pmatrix}  
\begin{pmatrix}
1& \ell\\
-K\ell & 1
\end{pmatrix}&\approx
  \begin{pmatrix}
1 & 0\\
-\eta/\rho & 1
\end{pmatrix}  
\begin{pmatrix}
1& \ell\\
-K\ell & 1
\end{pmatrix}
=
\begin{pmatrix}
1 &\ell\\
-\left(k_F\ell + \ell/\rho^2 + \eta/\rho\right) & -\eta\ell/\rho + 1
\end{pmatrix}
\\
&\cong
\begin{pmatrix}
1 &\ell\\
-\left(k_F\ell + \ell/\rho^2 + \eta/\rho\right) & 1
\end{pmatrix} =
\begin{pmatrix}
1 &\ell\\
-\left(k_F\ell + (\vartheta +\eta)/\rho\right) & 1
\end{pmatrix}\\
&\equiv
\begin{pmatrix}
1 & \ell\\
-1/f_F & 1
\end{pmatrix}
\end{aligned}
\label{eq::gp2}
\end{equation}
because $\ell⁄\rho^2 \cong \vartheta/\rho$, where $\vartheta$ is the sector bend angle and the length $\ell$ is the half magnet length for the quadrupole. The edge angle here has been assumed small to allow the tangent function to be approximated.

One can immediately see that the sector angle and edge angle term increase the focusing in the horizontal plane for a positive bend angle or dipole component.  Both the centripetal and edge angle term add constructively with the strong focusing.  Both planes are not identical, however, for in the vertical only the strong focusing and edge-angle terms contribute to a change in focusing strength.  In Eq.~\ref{eq::gp3} below the focal length of a FODO cell is expressed in terms of strong focusing, weak or body focusing, and edge focusing in the horizontal and vertical planes, respectively.  Note that the edge and body focusing are strongly dependent on the bending radius and become significantly stronger for compact rings. The different focusing terms can be varied independently to optimize machine parameters such as footprint, aperture, and tune. 
\begin{equation}
\left.
\begin{aligned}
\frac{1}{f_F} 
    &= k_F\ell + \frac{\vartheta +\eta}{\rho}\\
\frac{1}{f_D} &= k_D \ell + \frac{\eta}{\rho}
 \end{aligned}
 \qquad\right\}
  \label{eq::gp3}
\end{equation}

\subsubsection{Short straight inserts}
A short straight (SS) is a symmetric insertion formed in essence from a DFD cell with the middle F quad split into two to create a central drift space. The central drift can accommodate two RF modules. SS parameters are given in Table~\ref{tab::ss}.

\begin{table}
\centering
\begin{center}
\caption{Parameters of a short straight insert} 
\label{tab::ss}
\resizebox{\textwidth}{!}{
\begin{tabular}{|c |c|c|c|c|c|c|c|}
\hline
{\bf Device } & \bf Middle & \multicolumn{2}{c|}{ \bf QF} &\multicolumn{2}{c|}{\bf QD} &\multicolumn{2}{c|}{\bm{${\rm QD}_{\rm DS}-{\rm QD}_{\rm SS}$}}\\
& \bf drift & \multicolumn{2}{c|}{} &\multicolumn{2}{c|}{} &\multicolumn{2}{c|}{\bf interface}\\
\hline
\bf Param.& \bf length & \bf length& \bf gradient &\bf length& \bf gradient & \bf length & \bf gradient\\
\hline
\bf Value & 5.0000 m & 0.34302 m & 5.11671 T/m & 0.26317 m & 5.38490 T/m & 0.38833 m & 5.38490 T/m\\
\hline
\end{tabular}
}
\end{center}
\end{table}

\subsubsection{Long straight insertions}

The long straights (LS) are symmetric insertions created from a combination of quarter-wave transformers and doublet focusing. The length of the long central drift is adequate to house the injection stripping foil and associated ancillary orbit bump magnets. Parameters appear in Table~\ref{tab::ls}

\begin{table}
\centering
\begin{center}
\caption{Parameters of a long straight insert} 
\label{tab::ls}
\resizebox{\textwidth}{!}{
\begin{tabular}{|c |c|c|c|c|c|c|c|c|c|}
\hline
{\bf Device } & \bf Middle & \multicolumn{2}{c|}{ \bf QF} &\multicolumn{2}{c|}{\bf QD}& \multicolumn{2}{c|}{\bf QDX($\bm{\lambda/4}$)}&\multicolumn{2}{c|}{\bm{${\rm QD}_{\rm DS}-{\rm QD}_{\rm SS}$}}\\
& \bf drift & \multicolumn{2}{c|}{} &\multicolumn{2}{c|}{}&\multicolumn{2}{c|}{} &\multicolumn{2}{c|}{\bf interface}\\
\hline
\bf Param.& \bf length & \bf length& \bf gradient &\bf length& \bf gradient & \bf length & \bf gradient\bf gradient & \bf length & \bf gradient\\
\hline
\bf Value & 8.8575 m & 0.30195 m & 5.11671 T/m & 0.24701 m & 5.38490 T/m & 0.10048 m & 5.38490 T/m & 0.36470 m & 5.38490 T/m\\
\hline
\end{tabular}
}
\end{center}
\end{table}

\section{Injection}

\begin{wrapfigure}{C}{0.45\textwidth}
\centering
\includegraphics[width=0.45\textwidth]{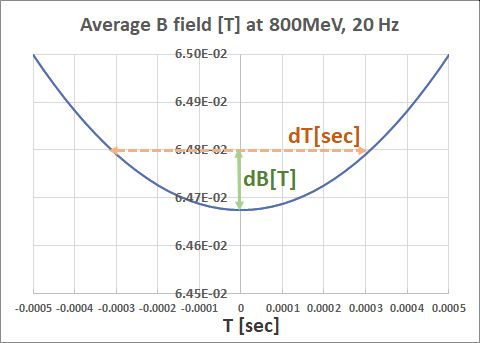}
\caption{Booster bend field for 0.8 GeV injection at a repetition rate of 20 Hz.}
\label{fig:flatramp}
\end{wrapfigure}

As can be seen in Fig.~\ref{fig:flatramp}, the bend field of the Booster is never flat. Therefore, a combination of incoming beam energy and bend field compensation with RF feedback manipulations have to be done to account for the long injection times of approximately $500$~$\mu$s from the PIP-II Linac. 

PAR alleviates this difficulty by accumulating beam before transferring it in a single turn injection to Booster. This simpler process is more controllable because of the reduced hardware complexity.
As noted above, one use of PAR will be to improve the loading of Booster. The RF phase lock process will be be much simpler because the PAR field is flat. In contrast, a rapid cycling synchrotron (RCS) like the Booster, uses resonant-circuit power supplies for the magnets and therefore has no true flat energy during injection or extraction. This makes loading any RCS especially challenging because the beam orbit can move horizontally by several millimetres if injection takes more than 10~$\mu$s.\cite{flat} 



\begin{wrapfigure}{C}{0.45\textwidth}
\centering
\includegraphics[width=0.45\textwidth]{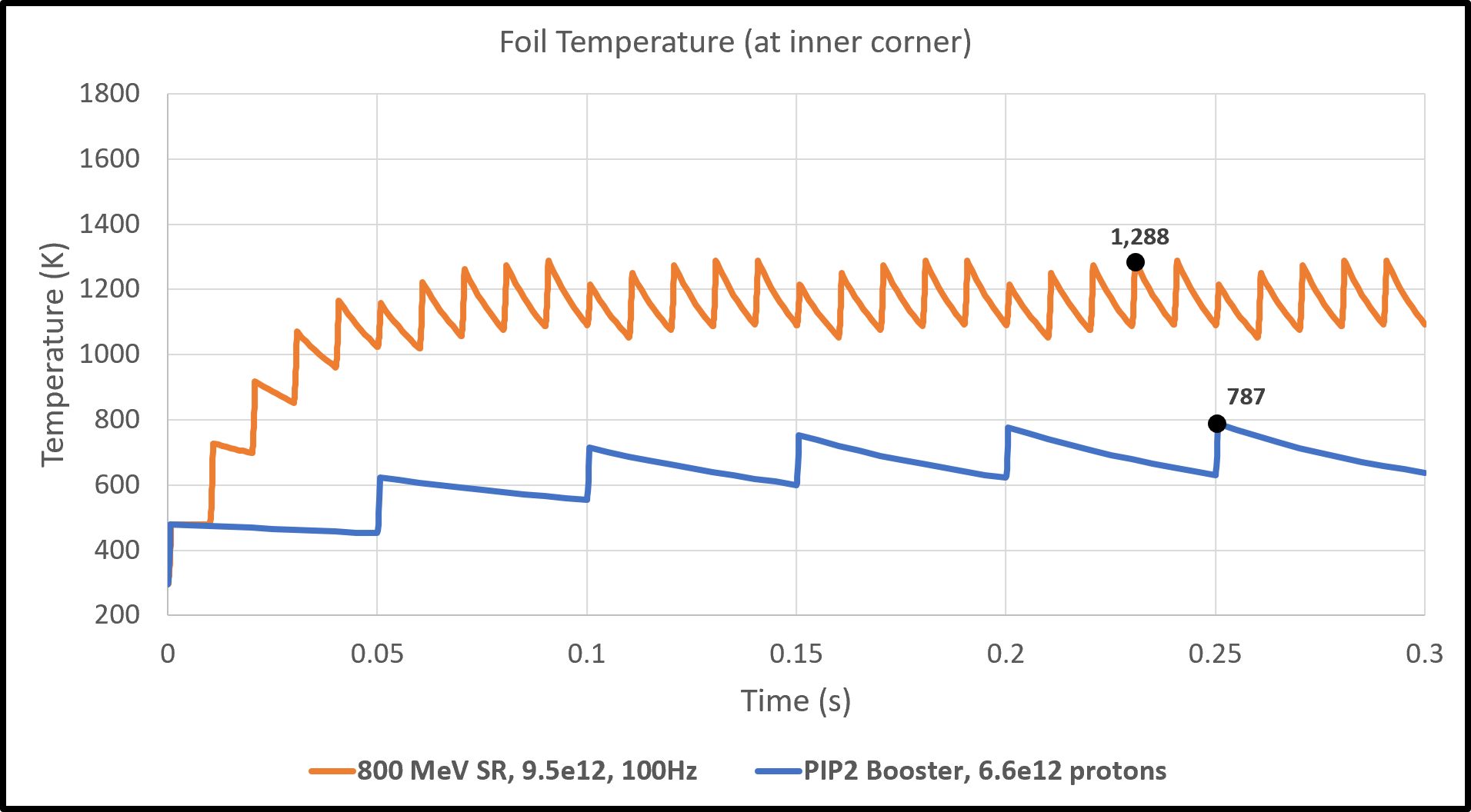}
\caption{Projected injection foil heating for 0.8 GeV PAR injection.}
\label{fig:foil}
\end{wrapfigure}

The foil injection constraints associated with 100~kW beam power delivery to the proposed 0.8~GeV beamline concurrent with 20~Hz delivery to Booster has been evaluated. The heating of the injection foil by the circulating beam was calculated using the method outlined in Ref.~\cite{Eldred}. Four out of five 100~Hz pulses use parameters for the 0.8~GeV physics program (9.5e12 protons and a 95\% normalized emittance of $24\pi$~mm~mrad) and the remaining one out five pulses use the nominal parameters for filling Booster (6.6e12 protons and emittance of $16\pi$~mm~mrad). The foil thickness of 600 $\mu$g/cm$^2$ and the product of $\beta_{x}$ and $\beta_{y}$ of 125~m$^{2}$ was retained from the Booster injection scenario outlined in the CDR~\cite{PIP2}. The result of the calculation is shown in Fig.~\ref{fig:foil} and compared to the scenario with Booster pulses only.

The hot spot of the injection foil reaches a peak temperature of $\sim
\!1300$~K, which is far below the temperature of 1800~K where foil sublimation begins to limit foil lifetime. Therefore,  no operational difficulties associated with an annually changed injection foil is anticipated.

The 100~kW pulsed proton beam for the 0.8~GeV physics program that is injected into PAR is a factor of 7 higher in power than the beam for injection into Booster. Consequently injection absorbers for the unstripped H$^{-}$ and H$^{0}$ ions, as well as circulating protons scattered off the injection foil will have to be optimized carefully.

PAR will support the following newly proposed 0.8 GeV HEP programs at Fermilab: 
\begin{enumerate}
\item  DS (dark energy sector) physics~\cite{DS1}, 
\item PRISM/PRIME type experiment~\cite{PRISM}, and 
\item a charged lepton flavor violation program.\cite{CLFVP}.  
\end{enumerate}

All these experiments demand 0.8 MeV intense short bunches of lengths in the range of 12 to 500~ns. The plan is to conduct systematic investigations on the possibility of using a 2.5 MHz (h=4) type RF system that is currently used in Recycler for $500$~ns bunches and to use Booster type RF systems for short bunches. Barrier bucket RF systems, that have been used in the past at Fermilab, can be used to produce compressed bunches of variable bunch lengths as needed by the experimenters. 

\section{Extraction}

\begin{wrapfigure}{C}{0.45\textwidth}
\centering
\includegraphics[width=0.45\textwidth]{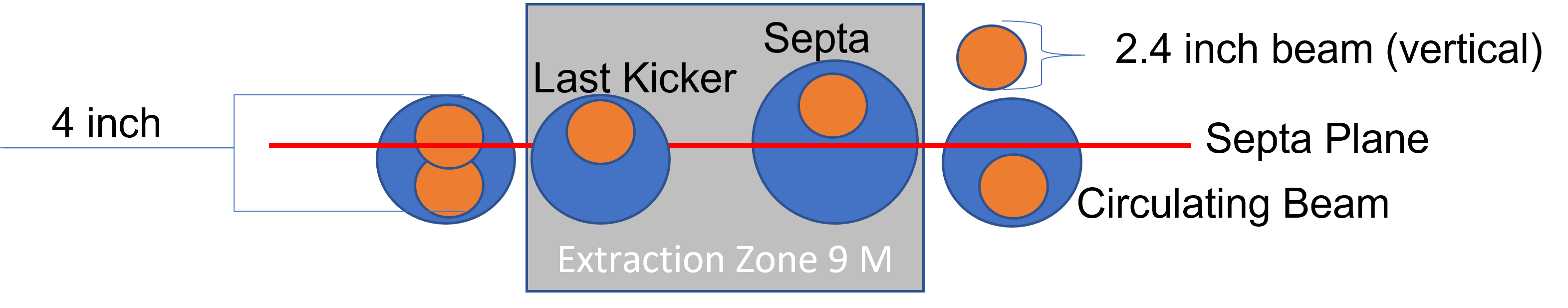}
\caption{The PAR vertical extraction region with the last kicker and septa shown.}
\label{fig:BAR2Ext}
\end{wrapfigure}

\begin{wrapfigure}{C}{0.45\textwidth}
\centering
\includegraphics[width=0.45\textwidth]{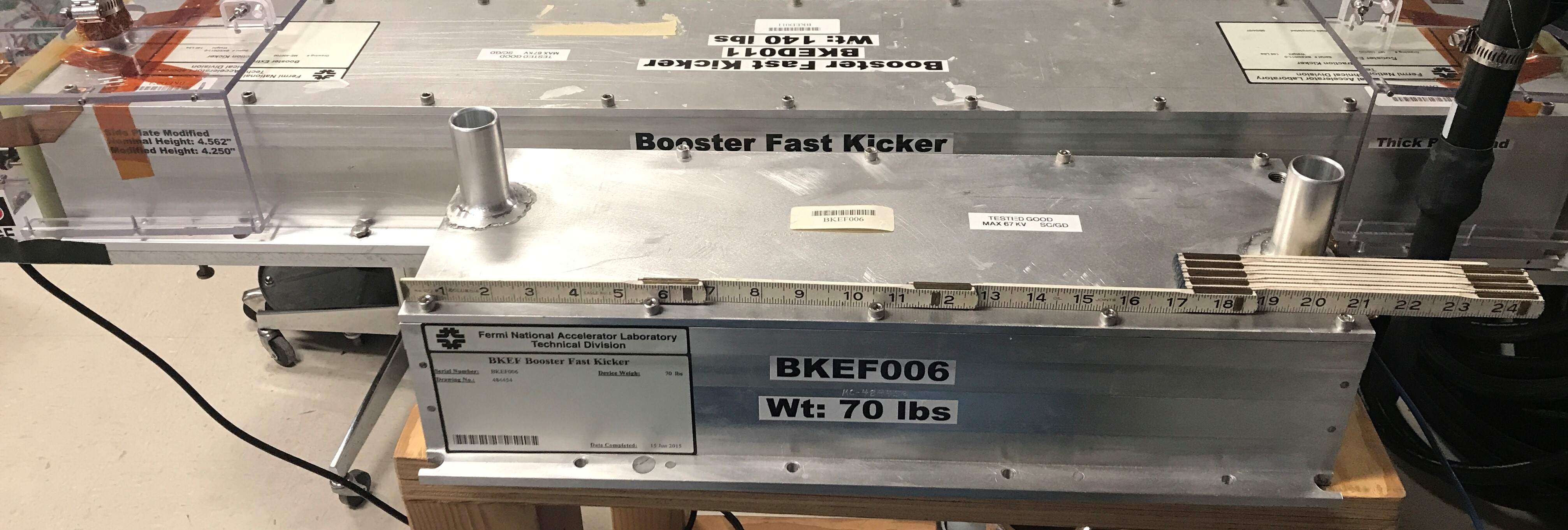}
\caption{An example of a short kicker for PAR extraction is the Booster style short kicker.}
\label{fig:Kicker}
\end{wrapfigure}

One of the critical factors of PAR for determining the beam delivery rate will be the extraction system.  High rate extraction systems do exist at other accelerator facilities but the PAR extraction goal of 100 Hz will be challenging. A conceptual vertical extraction of the beam from PAR to the BTL is shown in Fig.~\ref{fig:BAR2Ext}. The preliminary design concept is to enlarge the aperture of the existing half meter Booster 20 Hz kicker magnets to 3 inches to match the ring magnet apertures. See Fig.~\ref{fig:Kicker}. The cost of these kickers can be lowered by replacing their ceramic beam pipes with ribbed Inconel beam pipes.  Another cost saving solution that is being investigated, but is too early to base a design on, is to use PEEK.  The extraction region will be in the long straight near injection with a planned length of 8 to 9~m.  The length of the straight section will allow for extra kickers to be installed which reduces the required gradient. Or if there is enough space for 2 sets of kickers, then sequentially pulsing between them for extraction can be considered. For comparison, Booster has 8 kickers in a 4.5~m section for 8 GeV extraction while in PAR, extraction at 0.8~GeV will require half the number of kickers.

Extraction from PAR will likely occur in the same region as injection. One possible configuration, shown in Fig.~\ref{fig:Footprint}, is to have the injection line in the same plane as the upper PAR ring. The extraction line will be on the lower PAR ring. There will be dipole magnets in this line to bend the extracted beam to the same plane as the BTL. Note that there are multiple configurations which can accommodate the injection beam line, PAR rings and the extraction beam line.  A shifted PAR ring arrangement in both longitudinal and transverse space will minimize overlap issues.

\section{Users}

The quest to identify and understand dark matter (DM) is one of the most compelling missions in particle physics today.  Recent theoretical work has identified several classes of sub-GeV DM, which are neutral under the Standard Model (SM) but charged under new “dark sector” forces that mediate DM interactions with SM particles and which achieve the correct relic abundance in the early universe through a standard thermal-freeze out mechanism~\cite{Battaglieri:2017aum}.  What is most exciting about these classes of DM models is that, in the case that the mediator is heavier than the DM, they predict a SM-DM coupling as a function of DM mass that is tantalizingly close to current limits and can be probed by next-generation accelerator-based fixed target dark sector searches in the coming decade~\cite{Batell:2009di,deNiverville:2011it,deNiverville:2012ij,Gardner:2015wea,Berlin:2018pwi,Tsai:2019mtm,Berlin:2018bsc,Batell:2018fqo,Banerjee:2017hhz,Ariga:2019ufm,Alekhin:2015byh,NA62:2017rwk}.
\par Proton beam dump experiments are potentially sensitive to these and any other dark sector models that produce light DM directly through hadronic interactions or through the subsequent decay of light mesons. This includes, for example, both standard ``vector portal'' DM models that can be probed with both proton and electron beams as well as other models, such as hadrophilic DM, for which proton beams provide unique sensitivity~\cite{Batell:2018fqo}.  Recent results from the COHERENT~\cite{COHERENT:2021pvd} and Coherent CAPTAIN-Mills experiments~\cite{CCM:2021leg,Aguilar-Arevalo:2021sbh} have demonstrated how detectors capable of measuring coherent elastic neutrino-nucleus scattering (CEvNS) can also be used to set limits on vector portal and leptophobic DM at proton beam dumps.
\par
\begin{figure}
\centering
  \includegraphics[width=0.54\textwidth]{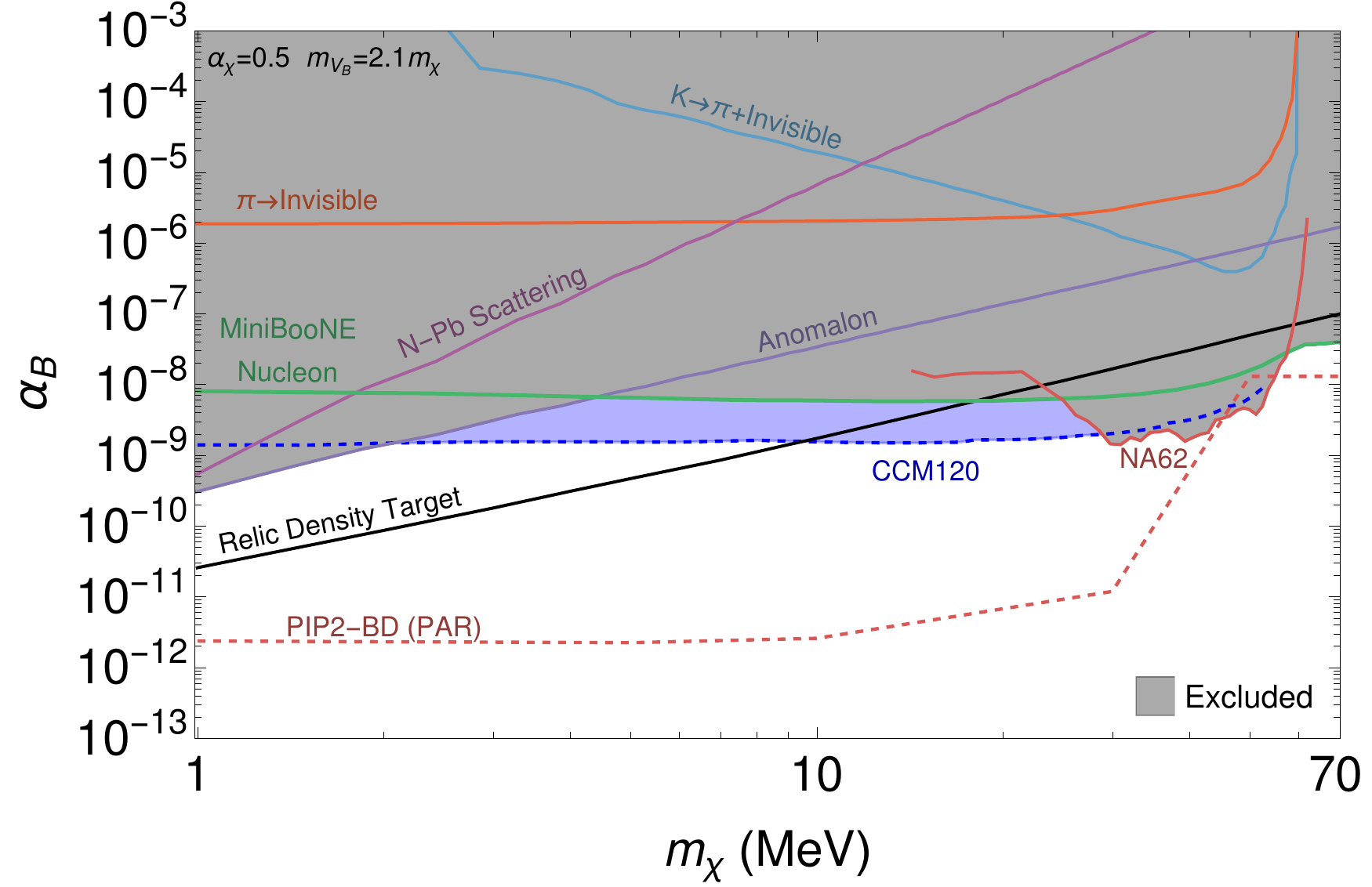}\hspace{1cm} \includegraphics[width=0.36\textwidth]{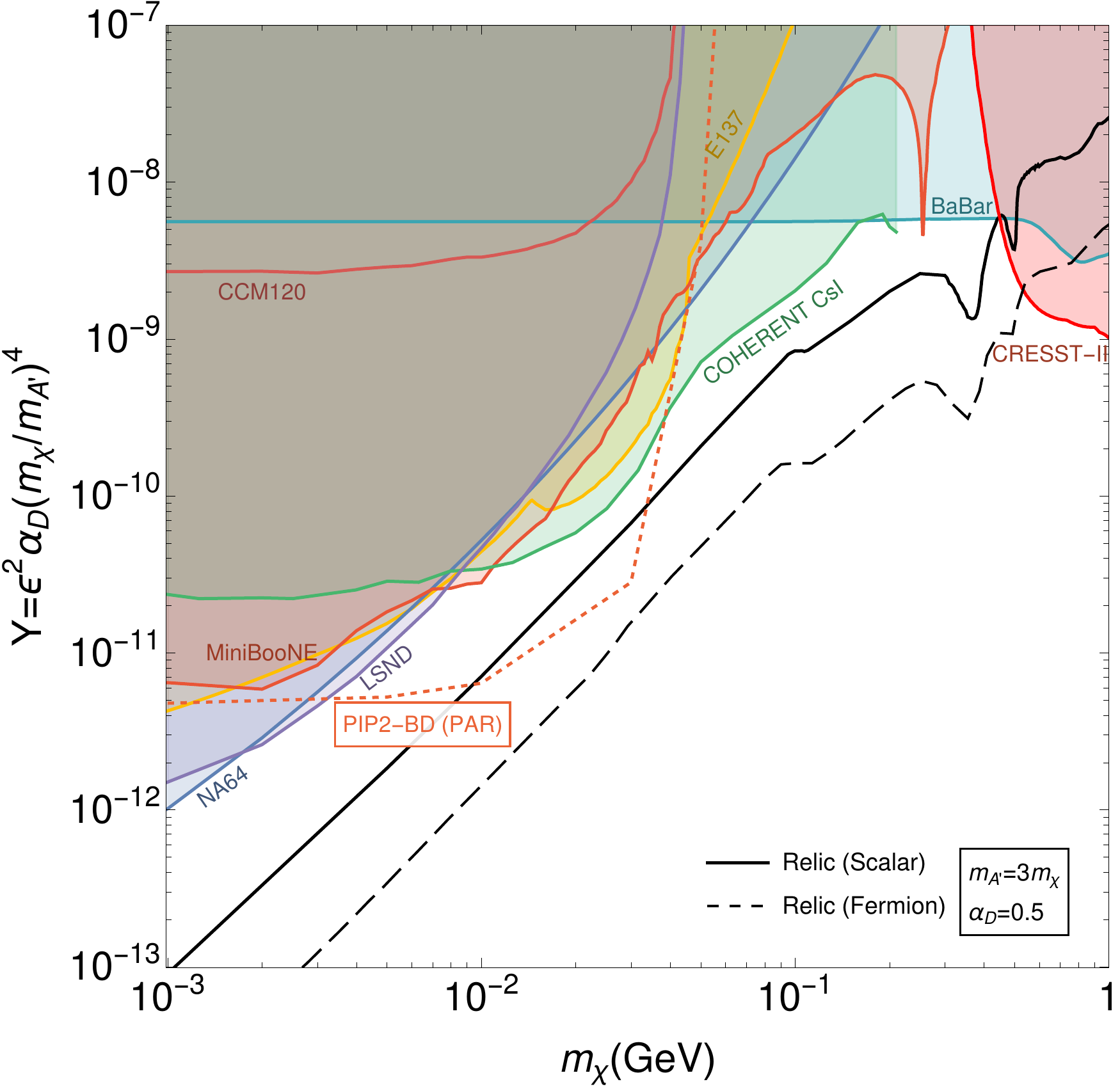}
  \caption{PIP2-BD DM 90\% C.L. sensitivity curves for $1.2\times10^{23}$ protons on target compared to thermal relic density targets and existing 90\% exclusion limits. Leptophobic DM limits (left) are shown in terms of the leptophobic coupling strength, $\alpha_B$, assuming $m_V=2.1m_{\chi}$ and $\alpha_\chi=0.5$ while vector portal DM limits (right) are shown in terms of the dimensionless scaling variable $Y=\epsilon^2\alpha(m_\chi/m_{A'})^4$, assuming $\alpha=0.5$ and $m_A=3m_{\chi}$.}
  \label{PIP2-BD_Senstivity}
\end{figure}

 We propose a 100-ton LAr scintillation-only detector, PIP2-BD~\cite{PIP2-BD}, with a design inspired by COHERENT and CCM, built within the decade and placed on-axis, 18 m downstream from a carbon proton beam dump driven by PIP-II and the PAR.  We use a Geant4-based target simulation and BdNMC~\cite{deNiverville:2016rqh} to generate both DM signal as well as CEvNS neutrino backgrounds in the detector.  The Geant4-based detector simulation includes instrumental effects (PMT dark pulses) as well as radiological backgrounds ($^{39}$Ar).  Applying a 50 keV nuclear recoil energy threshold (to suppress neutrino backgrounds) and a pulse-shape discrimination cut giving an efficiency of 70\%, we compute the 90\% C.L. sensitivity to a leptophobic and vector portal DM signal above the primarily CEvNS background using a $\Delta \chi^2$ test statistic, assuming a 5-year run and a 75\% uptime, as shown in Fig.~\ref{PIP2-BD_Senstivity}.  With this setup, we are able to set world-leading limits on leptophobic and vector portal DM and probe thermal relic density targets for scalar DM.
\\

\section{Conclusion}
The implementation of PAR offers the following significant benefits for the Fermilab HEP program: 
\begin{itemize}[noitemsep]
  \item Improved injection into Booster for PIP-II with 
  a robust beam dump. This relieves the injection front porch challenges in Booster and gives an option for increasing the beam intensity injected into Booster beyond the PIP-II design goals. 
    \item Extremely cost effective because the folded ring only requires half the Booster circumference and it also uses part of the BTL enclosure.
    \item Ability to adapt to changes in PIP-II energy with a simplified RF phase locking of the two machines. 
    \item Reduced PIP-II - DUNE power ramp up time. 
    \item A fast track to FNAL Dark Sector physics, PRISM type experimental and charged lepton violation programs with flexible bunch structures.
    \item Significant utilization of PIP-II capabilities from day 1.
\end{itemize}

For a comparison, Table~\ref{tab::basicparams} summarizes the current and future parameters of the accelerator systems when PAR is included.

\cellcolor[gray]{1}
\begin{table}[htbp]
\centering
\begin{center}
\caption[Milestones]{Accelerator Systems and Basic Parameters (for reference).} 
\label{tab::basicparams}
\begin{tabular}{|p{2.8cm}|c |c |c|c|}
\hline
{\bf Accelerator $^*$} & \bf Rep Rate Hz & \bf Beam Power MW & \bf Injection E GeV & \bf Extraction E  \\
\hline
Booster & 15 -- 20 & 0.089 -- 0.150 & 0.4 -- 0.8 & 8.0\\
\hline
 PIP-II & 20 -- CW &  2 & 0.0012 & 0.8 -- 1.0 \\
\hline
PAR & 20 -- 100  & 0.1 -- 0.2 & 0.8 -- 1.0 & 0.8 -- 1.0 \\
\hline
\hline
\end{tabular} 
\footnotesize{*This list includes the new storage ring, PAR, whose power output is not fully assessed. \\ 
 }
\label{tab:InhibitedCycles}
\end{center}
\end{table}

\section{Acknowledgement}
This manuscript has been authored by Fermi Research Alliance, LLC under Contract No. DE-AC02-07CH11359 with the U.S. Department of Energy, Office of Science, Office of High Energy Physics.

\bibliographystyle{unsrt}
\bibliography{references}
\end{document}